\documentclass[iop,apj]{emulateapj}

\usepackage{natbib}
\usepackage{graphicx}	
\usepackage{amsmath}	
\usepackage{amssymb}	
\usepackage{units}
\usepackage{enumitem}
\usepackage{bm}
\usepackage{color}
\usepackage[breaklinks,colorlinks,citecolor=blue,linkcolor=red]{hyperref} 

\emergencystretch=\maxdimen
\newcommand{\be}{\begin{equation}}
\newcommand{\ee}{\end{equation}}

\newcommand{\dd}{{\rm d}}

\def\h2{${\rm\,H_2}$}

\newcommand{\beq}{\begin{equation}}
\newcommand{\beqa}{\begin{eqnarray}}
		 \newcommand{\eeq}{\end{equation}}
\newcommand{\eeqa}{\end{eqnarray}}

\definecolor{orange}{rgb}{1, 0.5, 0}

\begin{document}

\title{A new approach to constrain the Hubble expansion rate at high redshifts by gravitational waves}


\author{Mohammadtaher Safarzadeh\altaffilmark{1}, Karan Jani\altaffilmark{2}, Nianyi Chen\altaffilmark{3}, Tiziana DiMatteo\altaffilmark{3,4}, Abraham Loeb\altaffilmark{5}}
\affil{$^1$Gravitational Astrophysics Laboratory, NASA Goddard Space Flight Center, Greenbelt, MD 20771, USA; \href{mailto:mtsafarzadeh@gmail.com}{mohammadtaher.safarzadeh@nasa.gov}}
\affil{$^2$Department Physics \& Astronomy, Vanderbilt University, 2301 Vanderbilt Place, Nashville, TN, 37235, USA}
\affil{$^3$McWilliams Center for Cosmology, Department of Physics, Carnegie Mellon University, Pittsburgh, PA 15213}
\affil{$^4$NSF AI Planning Institute for Physics of the Future, Carnegie Mellon University, Pittsburgh, PA 15213, USA}
\affil{$^5$Center for Astrophysics | Harvard \& Smithsonian, 60 Garden Street, Cambridge, MA}

\begin{abstract}
Detection of massive binary black hole (BBH) mergers at high redshifts is a target for LISA space mission. While the individual masses of a BBH merger are redshifted, the mass ratio of BBH mergers is independent of their redshift. Therefore, if there is an independent correlation between the mass ratio and redshift, such a relationship can be used to i) infer the redshift of the merging binaries, and together with the luminosity distance measurement ($D_L$), constrain the expansion rate of the universe at high redshifts $H(z)$, and ii) constrain models of supermassive black hole seed formation in the universe assuming a fixed cosmology. We discuss why there is an expected relation between the mass ratio of the massive BBHs with their redshift and show the clues for this relation by analyzing cosmological hydrodynamical simulations of BBH mergers. This approach opens up the possibility of directly measuring the expansion rate at redshift $z \approx 2$ with LISA for the first time. Moreover, we discover a trend between seed mass and mass ratio of massive BBHs which by itself is a major result that could be exploited to constrain the formation scenarios of supermassive BH seeds.
\end{abstract}


\section{Introduction}
\label{sec:intro}

Measurement of the Hubble expansion rate of the universe is of utmost importance to our understanding of cosmology, such as the nature of dark energy and modified gravity \citep[e.g., ][]{FTH2008,Clifton2012PhR}. In the era of gravitational wave (GW) astronomy, the standard sirens have provided a novel avenue for measuring the local expansion rate ($H_0$). The first of such measurements was carried out by identifying the host galaxy of the binary neutron star merger event GW170817 \citep{Abbott2017GW170817} leading to an estimate of $H_0= 70.0^{+12.0}_{-8.0}\rm kms^{-1}Mpc^{-1}$. While there has been a proposal to measure the expansion rate at higher redshifts using features in the mass distribution of the LIGO binary black holes \citep{Farr2019ApJ,EH2022PhRvL}, measuring the expansion rate at $z\approx2$ is only inferred from Baryonic Acoustic Oscillations (BAO) signatures in Ly-$\alpha$ forest by assuming radiation drag is known through Cosmic Microwave Background (CMB) or BigBang Nucleosynthesis (BBN) \citep{Font-Ribera2014JCAP,Delubac2015,Bautista2017,Agathe2019}. 

In this work we propose a novel approach where direct measurement of the expansion rate of the universe at high reshifts $z\approx2$ is possible through the gravitational wave signal from massive BBH mergers observable with next-generation space-based missions, Laser Interferometer Space Antenna \citep[LISA, ][]{LISA2017,LISA2019} which can study the coalescence of BHs in the mass range of $10^4-10^7 M_{\odot}$ and redshifts below $z<20$.

While the individual mass components of a merging MBBHs are redshifted as $m_{\rm obs}=m_{\rm source}(1+z)$, where $m_{\rm obs}$ and $m_{\rm source}$ are observed and source mass respectively, the mass ratio of an MBBH is independent of redshift. Therefore, if there is an independent relation between mass ratio and redshift, it becomes possible to statistically infer the redshift of a merging MBBH source from its mass ratio, $P(z_i\mid q_i)$. Combining this with the inferred luminosity distance to the source ($D_{L_{i}}$) from LISA GW observation, the expansion rate of the universe at the redshift of the source can be inferred. An ensemble of observations of MBBHs can then constrain the high redshift Hubble expansion rate as $P(H(z)\mid D_L,z)$ to high accuracy. The way to get $H(z\approx2)$ is through measuring the slope of the change in $D_L$ and redshift:

\begin{equation}
\frac{\partial D_L}{\partial z} = \frac{D_L}{(1+z)} +\frac{(1+z)c}{H(z)}\,, 
\end{equation}
with the luminosity distance defined as 
\begin{equation}
    D_L=(1+z)D_H\int_0^z\frac{\dd z'}{E(z')}\,,
\end{equation}
where $c$ is the speed of the light, $D_H=c/H_0$ is the horizon distance, and $E(z)=H(z)/H_0$. Since the hydrodynamical simulations have shown the peak of the merger rate of MBBHs to be around $z\approx2$ \citep{Ni2022}, our proposed technique will constrian the Hubble expansion rate at similar redshifts.

Why do we expect MBBHs to have a relationship between their merging redshift and their mass ratio? The formation of supermassive black holes is a matter of debate \citep{Smith2019,Volonteri2021} with two main possible formation scenarios: i) they form through direct collapse \citep{Bromm2003} or ii) they form through accretion onto small stellar seed black holes \citep{Madau2001}. Regardless of what is the true nature of these massive BHs, one expects that merging MBBHs have more similar masses in the early universe with their mass ratio starting to deviate from equality at lower redshifts, due to hierarchical structure formation in the universe. This would mean that the probability distribution function (PDF) of mass ratio for merging MBBHs at high redshifts peaked around $q\approx 1$ (assuming the seed mass is the same for all the BHs), and this PDF is expected to widen and skew towards lower values at lower redshifts. 

Our proposed approach works if the GW probe's bias in the mass ratio is not severe and can be corrected for. We show in Figure 1 that while LISA's observable volume is sensitive towards large mass ratios, in this work our result is based on a mass ratio cut of about 0.5 which we later show can be easily accounted for.

One of the main advantages that our method provides over most other methods suggested in the literature to constrain Hubble expansion rate with standard sirens is that we do not rely on the electromagnetic counterpart of the gravitational wave sources and therefore we do not need to localize the source to high accuracies which are the bottleneck facing the GW probes \citep{Tamanini2016JCAP}. Moreover, this technique \emph{directly} measures the expansion rate at $z \approx 2$ where currenlty is only indirectly inferred by observing BAO feature in $\rm Ly-{\alpha}$ forest \citep{Font-Ribera2014JCAP,Delubac2015,Bautista2017,Agathe2019}, therefore providing a complementary probe that can be tested against the BAO measurements.

In \S\ref{sec:method} we verify our assumption regarding the existence of a correlation between mass ratio and redshift by analyzing the result of cosmological hydrodynamical simulations of merging MBBHs. In \S\ref{sec:result} we show our results and provide parametrized fits for the correlations we discover, and in \S\ref{sec:discussion} we discuss and provide caveats for using our approach to constrain the expansion rate of the universe at $z\approx2$.

\begin{figure}[ht!]
\epsscale{0.80}
\includegraphics[width=.95\columnwidth, trim = {0 -20 0 0}]{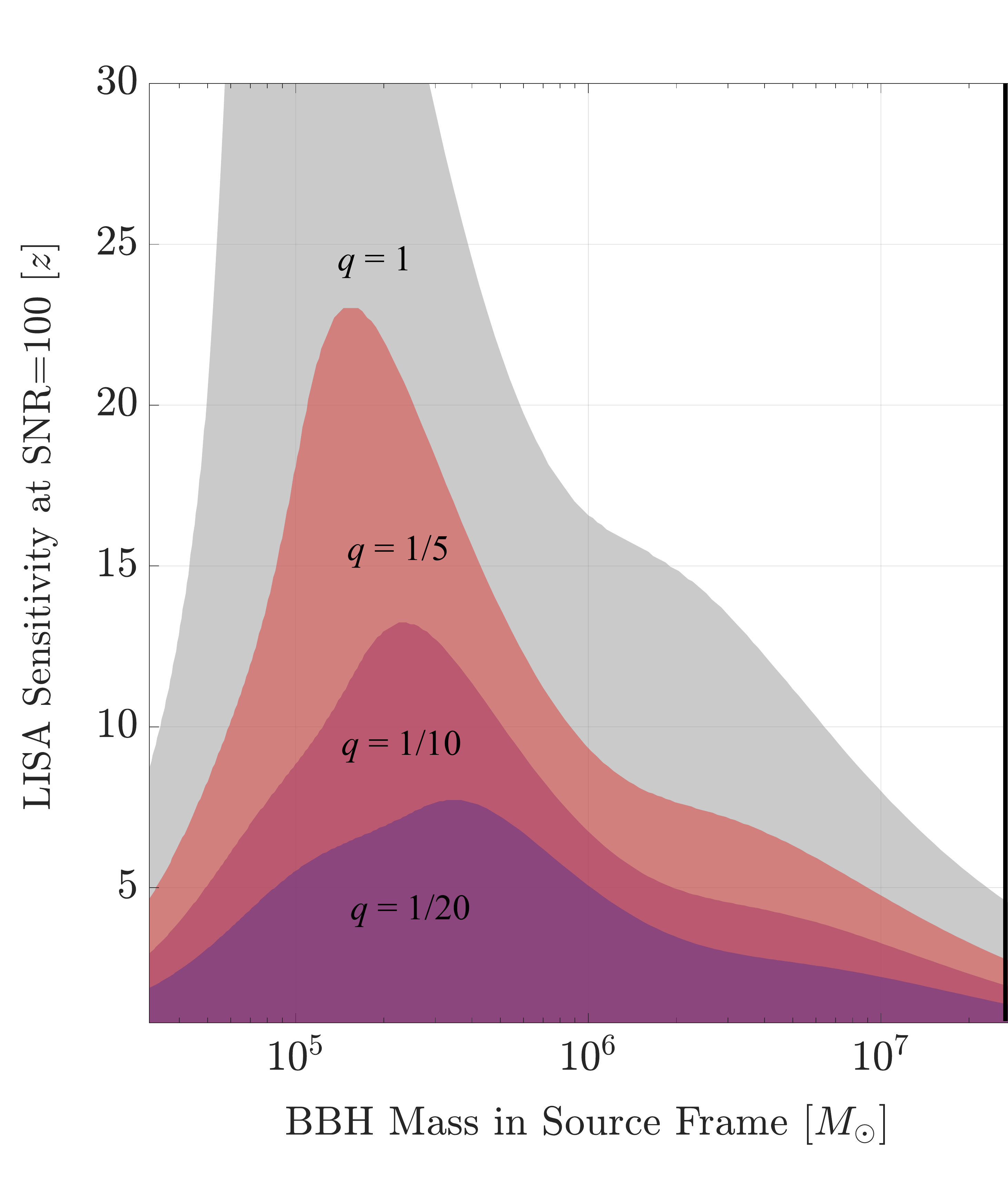}
\caption{The detection sensitivity of LISA, as measured in cosmological redshift, for BBH mergers of different mass and mass ratios. We compute the detection radius at a signal-to-noise ratio of 100. LISA can see equal mass ratio merger in the entire universe at around $10^5M_{\odot}$ while the observable redshift for the same mass bin drops to about $z\approx 12$ at mass ratios around $q\approx 0.1$.}
\label{fig:fig1}
\end{figure}

\section{Method}\label{sec:method}

We verify the assumption that the mass ratio of the merging MBBHs evolves with redshift by analyzing the very recent large-volume, high-resolution (gravitational softening of $1.5\,{\rm kpc}/h$; dark matter mass resolution of $9.6\times 10^6\,M_\odot$) cosmological hydrodynamic simulation \texttt{ASTRID}  \cite{Bird2022,Ni2022}.
\texttt{ASTRID} includes subgrid models for star formation, BH accretion, and the associated supernova and AGN feedback, recently updated with an SMBH seed population between $3\times 10^4\,M_\odot/h$ and $3\times 10^5\,M_\odot/h$ and a sub-grid dynamical friction model to follow the SMBH dynamics down to kpc scales.
With a comoving volume of $(250\,{\rm Mpc}/h)^3$, \texttt{ASTRID} is the largest galaxy formation simulation up to date that covers the epoch of the cosmic noon. 
The BHs are merged in the simulation when they are separated by $\Delta r<3\,{\rm ckpc/h}$ and are gravitationally bound to their local structure.
The range of BH seed mass down to $3\times 10^4\,M_\odot/h$ allows \texttt{Astrid} to capture the large population of high-redshift seed-seed mergers, as well as the SMBH-seed mergers down to $q<10^{-4}$.
The details of the MBH merger population in \texttt{ASTRID} can be found in \citet{Chen2022a}.

\begin{figure}[hb!]
\epsscale{0.80}
\centering
\includegraphics[width=1\columnwidth, trim = {150 0 0 -80}]{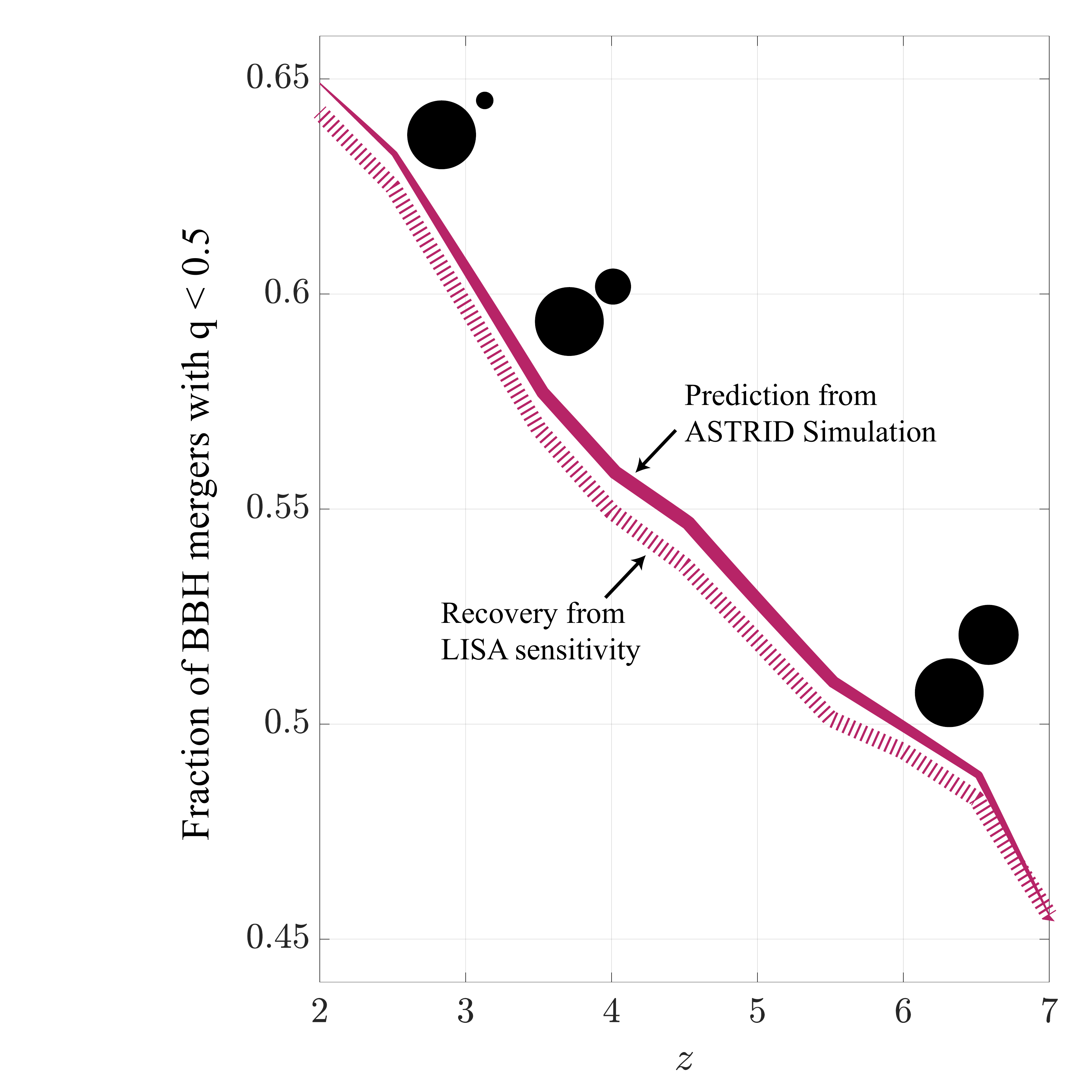}
\caption{The fraction of mergers with $q<0.5$ as a function of redshift ($z$) shown in solid purple. The dashed line shows how well LISA can recover the underlying trend despite the selection effects for sources with SNR $100$. The fact that LISA's selection effects does not affect the recovery of the true distribution (solid line) is promising. The entire dataset from \texttt{ASTRID} cosmological hydrodynamical simulations is used in this plot. We note that the curves end at $z\approx2$ which is the redshift that \texttt{ASTRID} simulation has been carried out to.}
\label{fig:fig2}
\end{figure}

We then convolve the true relation of the redshift-mass ratio through the LISA noise curve to estimate how faithful the recovery of the relationship would be.
To do so we first compute the signal-to-noise ratio of LISA sensitivity as prescribed in \cite{Robson_2019}. For computing the horizon distance (detection reach) of LISA at a given BBH masses in their source frame, we use the infrastructure described in \cite{2020NatAs...4..260J}. 

\section{Results}\label{sec:result}

Fig. \ref{fig:fig1} shows the observable universe of MBBHs for LISA as a function of BH seed mass categorized into different mass ratio bins. 
LISA can see equal mass ratios to much higher than it can for unequal mass ratio mergers. However, in this work, we are concerned with systems between the mass ratios of 1/2 and 1 which are not much impacted by the selection effects. If we had wanted to explore the case for more extreme mass ratios, then the selection effects of LISA would have played a more major role in analyzing the results.

Fig. \ref{fig:fig2} shows the observed relation between the mass ratio of the merging MBBHs and redshift. 
These results are from the \texttt{ASTRID} cosmological hydrodynamical simulations as described in the previous section.
The solid line shows the results from the simulations, and the dashed line shows the observed relation by LISA knowing the noise curve of the LISA as discussed in \citet{2020NatAs...4..260J}.
The close proximity of the dashed line to the solid line shows we can recover the true relation of mass ratio and redshift with LISA which makes this approach promising. 

We provide a power-law fitting formula for the results in Figure 2 as follows:
\begin{equation}
    f_{\mathrm{BBH}(q<0.5)} \propto z^{-1/5}
\end{equation}
\label{eq.1}

This power law can be used to statistically infer the redshift of the merging MBBHs and constrain the expansion rate of the universe by independently measuring the luminosity distance to the source, $D_L$ from GW observations.  

We emphasize that measuring the slope in Fig. 2 from observations assuming a fixed cosmology by itself is an important result which can constrain models of seed black hole formation in the universe. 

\begin{figure}[t!]
\epsscale{0.80}
\centering
\includegraphics[width=1\columnwidth, trim = {150 0 0 50}]{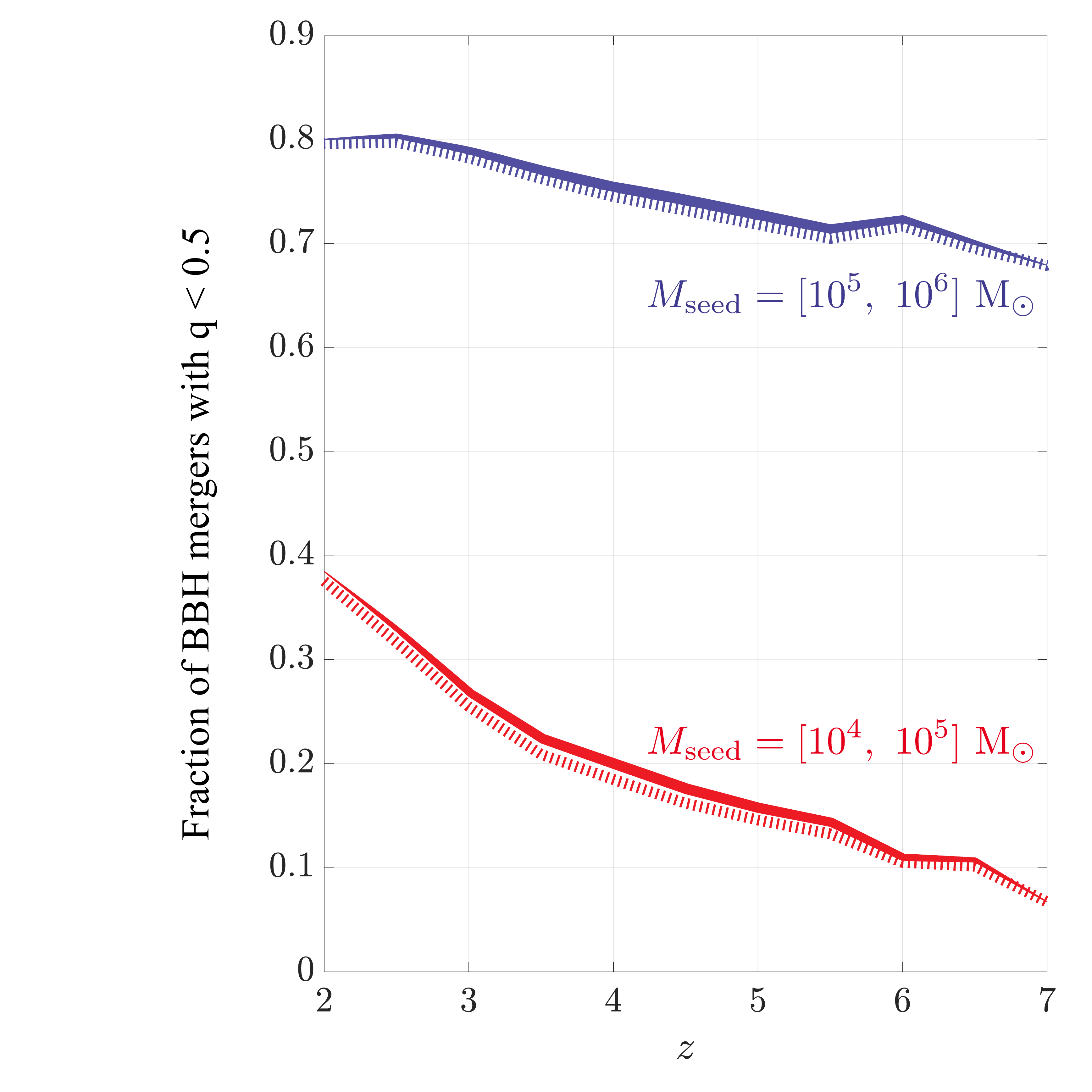}
\caption{The fraction of merger with $q<0.5$ as a function of z for different seed masses. The dashed line shows 
what LISA sees for sources with SNR $100$ for each seed mass respectively. We observe that at higher seed masses the trend becomes shallower. The fact that this trend is seed mass dependent can be used to constrain the formation scenario for the seed mass of supermassive BHs.}
\label{fig:fig3}
\end{figure}

Fig. \ref{fig:fig3} shows the same results discussed in Figure 2 but divides the data into different seed mass bins. We observe that the redshift-mass ratio relationship is sensitive to the seed mass of the BHs with higher seed masses showing a more flat relationship. 
This result on its own is interesting in that one can jointly fit the seed mass of the MBHs while measuring the expansion rate of the universe at high redshifts.
\begin{equation}
     f_{\mathrm{BBH}(q<0.5)} \propto  
    \begin{cases}
    z^{-1/7}~~~ \mathrm{for}~M_\mathrm{seed}=10^{5-6}~\mathrm{M}_\odot \\
    z^{-2/5}~~~ \mathrm{for}~M_\mathrm{seed}=10^{4-5}~\mathrm{M}_\odot
    \end{cases}
    \label{bounds}
\end{equation}

\section{Discussion and Implications}\label{sec:discussion}

The assumption that the mass ratio has an independent correlation with redshift, while expected from the arguments laid out in the introduction, needs to be verified by cosmological simulation assuming different cosmological parameters in order to make sure such a relation is not subject to a large variation assuming slightly different cosmologies. 
While we have not verified this assumption, we argue that given the large box size of the simulation, the results presented in this work do not heavily depend on the adopted cosmology and are more dependent on the hydrodynamical part of the simulations. 

The predicted merger rate of massive BBHs is about 0.3-0.7 mergers per year and can be expected to be more significant for more eccentric binaries. This would mean that about a couple of such systems are detected by LISA within its lifetime which is not enough to exploit the redshift-mass ratio of the merging binaries to constrain the expansion rate of the universe at high redshifts. 
However, the predicted merger rates are subject to uncertainties where it is possible for future cosmological hydrodynamical simulations to predict higher rates.
For example, in higher-resolution, smaller-box simulations such as Romulus25 \citep{Tremmel2017} and NewHorizon \citep{Volonteri2021}, it has been shown that resolving the MBHs in dwarf galaxies could increase the high-redshift merger rate prediction by an order of magnitude.

Gravitational-wave detector concepts beyond LISA, such as space missions targeting the deci-Hertz band \citep{2020CQGra..37u5011A, Izumi_2021}, milli-Hertz band~\citep{Baibhav_2021} and micro-Hertz band~\citep{Sesana_2021}, along with lunar detectors \citep{Jani_2021, Harms_2021} will access BBH mergers of masses from the distribution in Fig. \ref{fig:fig3}. The additional mission lifetime of these future detectors will increase the detection statistics to uncover the trends in eqs. \ref{eq.1}-\ref{bounds}.  

The advantage of this approach over other methods is the redshift range that is being probed. The peak of the merger rate for MBBHs is about $z\approx2$ and only handful of techniques can measure the expansion rate at such high redshifts\citep{Font-Ribera2014JCAP,Delubac2015,Bautista2017,Agathe2019}. For example, by detection of the BAO feature in the flux-correlation function of the $\rm Ly{\alpha}$ forest of high-redshift quasars, \cite{Delubac2015} measured  $H(z=2.34)=222\pm7(1\sigma)~\rm kms^{-1}Mpc^{-1}$. However, these techniques do not measure the expansion rate directly at at $z\approx 2$ since they rely on knowing the drag radius which is obtained through CMB or BBN. The techinque proposed in this work is able to measure the expansion rate directly and therefore be tested against the BAO approach and further constrain the expansion rate at $z \approx 2$.

\acknowledgements 
We are thankful to Licia Verde, Daniel Holz, and Eric Thrane for useful discussions. MTS's research was supported by an appointment to the NASA Postdoctoral Program at the (NASA Goddard Space Flight Center), administered by Oak Ridge Associated Universities under contract with NASA.

\bibliographystyle{yahapj}
\bibliography{ms}
\end{document}